\documentclass{article}
\usepackage{spconf,graphicx}
\usepackage{cite}
\usepackage{amssymb,amsfonts}
\usepackage{graphicx}
\usepackage{textcomp}
\usepackage{xcolor}
\usepackage{algcompatible}
\usepackage{soul}
\usepackage[ruled,vlined]{algorithm2e}

\usepackage{xurl}
\usepackage{booktabs}
\usepackage[ruled]{algorithm2e}
\usepackage{algpseudocode}
\usepackage{enumerate}
\usepackage[english]{babel}
\usepackage{blindtext}
\usepackage[caption=false,font=footnotesize,subrefformat=parens]{subfig}
\usepackage{amsmath}

\usepackage{xspace}
\usepackage{multirow}
\usepackage{threeparttable}
\usepackage{float}
\usepackage{flafter}
\usepackage{comment}
\usepackage[skip=12pt]{caption}
\usepackage{graphicx}
\usepackage{mathtools}
\usepackage[hidelinks,colorlinks=true,linkcolor=blue,citecolor=green,urlcolor=red,hyperfootnotes=false]{hyperref}

\def\BibTeX{{\rm B\kern-.05em{\sc i\kern-.025em b}\kern-.08em
    T\kern-.1667em\lower.7ex\hbox{E}\kern-.125emX}}

\def\fig{Fig.\xspace}
\def\eqn{Eqn.\xspace}

\def\tab{Tab.\xspace}

\def\ie{{\textit{i.e.}\xspace}} 
\def\eg{{\textit{e.g.}\xspace}}

\def\etc{{\textit{etc.}\xspace}}

\graphicspath{{./figures/}}
\graphicspath{{./pdf/}}
\DeclareGraphicsExtensions{.pdf,.jpeg,.png}

\ifodd 1
\newcommand{\com}[1]{\textbf{\color{red}(COMMENT: #1)}} %
\newcommand{\todo}[1]{\textbf{{\color{orange}(TODO: #1)}}}
\else

\newcommand{\com}[1]{}
\newcommand{\todo}[1]{}
\fi

\newcommand{\multiline}[1]{%
    \begin{tabularx}{\dimexpr\linewidth-\ALG@thistlm}[t]{@{}X@{}}
        #1
    \end{tabularx}
}

\title{Unfolding Target Detection with State Space Model}
\name{Luca Jiang-Tao Yu, Chenshu Wu}
\address{
The University of Hong Kong, Hong Kong 
}
\begin{document}
\maketitle
\begin{abstract}
Target detection is a fundamental task in radar sensing, serving as the precursor to any further processing for various applications. 
Numerous detection algorithms have been proposed. 
Classical methods based on signal processing, \eg, the most widely used CFAR, are challenging to tune and sensitive to environmental conditions. 
Deep learning-based methods can be more accurate and robust, yet usually lack interpretability and physical relevance. 
In this paper, we introduce a novel method that combines signal processing and deep learning by unfolding the CFAR detector with a state space model architecture. 
By reserving the CFAR pipeline yet turning its sophisticated configurations into trainable parameters, our method achieves high detection performance without manual parameter tuning, while preserving model interpretability. 
We implement a lightweight model of only 260K parameters and conduct real-world experiments for human target detection using FMCW radars. The results highlight the remarkable performance of the proposed method, outperforming CFAR and its variants by 10$\times$ in detection rate and false alarm rate. 
Our code is open-sourced here: \href{https://github.com/aiot-lab/NeuroDet}{\texttt{https://github.com/aiot-lab/NeuroDet}}.

\end{abstract}

\section{Introduction}
\label{sec:intro}

\begin{figure}[t]
    \centering
    \subfloat[CFAR pipeline.]{
    \label{subfig:trad_cfar}
      \includegraphics[width=0.9\linewidth]{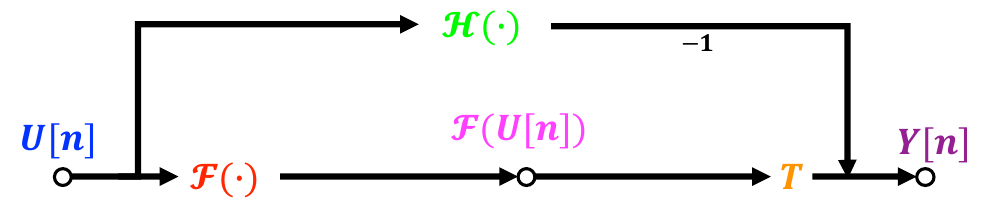}}
   \quad
    \subfloat[Continuous model pipeline of the state space model. \rm{The $\mathbf{dt}$ means the derivation of continuous step.}]{
    \label{subfig:lss_cfar}
      \includegraphics[width=0.9\linewidth]{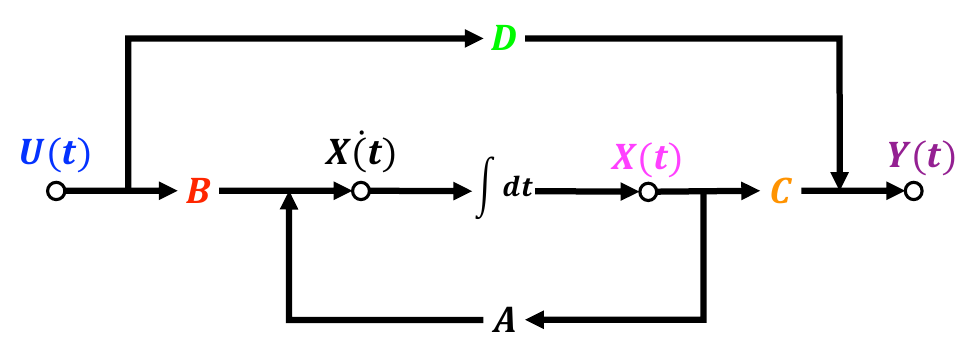}
   }
    \caption{Comparison of CFAR and continuous state space model. \rm{Colors show the correspondence relationship between them.}}
    \label{fig:comparison}
    \vspace{-0.1in}
\end{figure}

Detection is commonly recognized as the holy grail and a prelude task for most RF sensing applications, such as localization and tracking. So far, various detection algorithms have been proposed, \eg, improved matched filter \cite{villeneuve1999improved}, Bayesian detector \cite{ureten1999bayesian}, \etc, but most of them suffer from various limitations. For example, the matched filter requires knowing a reference signal a priori, which is frequently not available, \eg, in the presence of non-cooperative targets. The Bayesian detector also requires priority distributions and is sensitive to the environment. 

The CFAR detector is widely used for its adaptive thresholding and clutter suppression, maintaining a constant false alarm rate without requiring special signal distribution or prior knowledge. It evaluates the noise level of neighboring cells around the cell under test (CUT) to determine the threshold for detection. 
Among its many variants, Cell Averaging CFAR (CA-CFAR) is simple and computationally efficient but can be sensitive to non-uniform noise distributions. To address this, Ordered Statistics CFAR (OS-CFAR) \cite{gandhi1989adaptive} is introduced to improve performance in heterogeneous environments, yet it increases computation due to the ordering operation. Other variants, like the Greatest Of CFAR (GO-CFAR) \cite{hansen1973constant} and the Smallest Of CFAR (SO-CFAR) \cite{trunk1974modified}, aim to improve the detection in different scenarios. More advanced methods, such as adaptive \cite{bandiera2007adaptive, khalighi2000adaptive} and variation index CFAR \cite{zhu2021robust}, combine multiple measurement techniques at the cost of higher computational complexity. 
Additionally, these approaches all require careful parameter selection with extensive domain knowledge, and an improper configuration may lead to degraded performance.

In response to these drawbacks, machine learning (ML) techniques have been introduced to improve CFAR detection. Some researchers replace the measurement of CFAR with a Support Vector Machine (SVM) to select the best scheme \cite{wang2017intelligent}. However, the function of the SVM selector is limited because it can only choose the variants between CA-CFAR and GO-CFAR. Supervised deep learning is exploited in \cite{akhtar2018neural} and \cite{lin2019dl} to enable better detection. A recent work CFARnet \cite{diskin2024cfarnet} proves that CFAR can be maintained within a deep learning framework while providing computational efficiency and flexibility. 
Importantly, although these deep learning-based methods can improve the target detection performance, by incorporating prior knowledge in the trained models, they mostly leverage black-box neural networks and lack interpretability, raising concerns in RF sensing applications that are tightly coupled with the physical environment.

In this paper, inspired by recent advances in neural architectures based on state space models \cite{gu2021combining}, we present a novel signal processing-guided deep learning design that features the advantages of both approaches while overcoming their respective drawbacks. 
Specifically, our method unfolds the CFAR algorithm by devising a trainable network architecture that strictly follows the CFAR processing pipeline. 
We conduct experiments on CNNs, RNNs, and CFAR variants.
The results show that our method achieves approximately 10$\times$ higher detection rate at the same false alarm rate and 10$\times$ lower false alarm rate at the same detection rate than traditional CFAR variants while demonstrating significantly lower complexity compared to previous neural detectors based on CNNs and RNNs. 
Our approach also shows strong generalization ability to unseen datasets, thanks to the interpretable architecture.

\section{Methodology}
\label{sec:pf}

\subsection{CFAR Pipeline}

\label{sec:tcfar}
Denote the input sequence as $U[n]$, the CUT as $\bar{u}$, and the adjacent guard cells $u_g$. The sequences will pass by two different functions $\mathcal{F}(\cdot)$ and $\mathcal{H}(\cdot)$. The former works as \textit{selecting operator}, varying depending on the categories of CFAR, like average for CA-CFAR and selection after sorting for OS-CFAR. The later performs as \textit{testing operator}, normally a linear function, $\mathcal{H}(U[n]) = \bar{u}$. Then, the selected value $\mathcal{F}(U[n])$ will multiply the pre-defined threshold $\mathbf{T}$ (according to the constant false alarm rate), and subtract the tested value $\mathcal{H}(U[n])$ to get the output:
\begin{equation}
    \label{eqn:cfar}
    Y[n] = \mathbf{T} \cdot \mathcal{F}(U[n])-\mathcal{H}(U[n]).
\end{equation}

The output $Y[n]$ can be regarded as a sequence with the same length as the input after the sliding window, with the binary values representing the detection results for each cell. 
Albeit being efficient and scalable, the performance of CFAR detectors is sensitive to pre-configurations, \eg, selecting and testing operators, false alarm rate, and fixed sliding window size. 
Optimizing these parameter settings requires comprehensive domain knowledge of radar signals, and improper configurations may significantly degrade the performance.

\subsection{Network Model}

Our design is inspired by the linear state space model from control systems and recent advances in trainable linear state space models \cite{gu2021combining}. 
By incorporating activation functions into the trainable linear state space model, we can introduce non-linearity into the model. Consequently, this allows for the design of an interpretable unfolding detector based on the state space model.

\begin{figure}[t]
    \centering
    \includegraphics[width=0.9\linewidth]{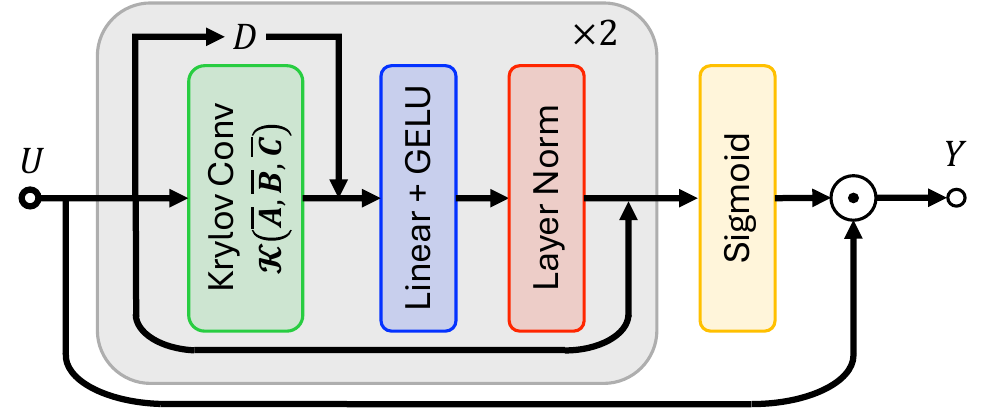}
    \caption{Model architecture.}
    \label{fig:model}
\end{figure}

The linear state space model maps the input sequence to the output sequence $U(t) \mapsto Y(t)$ via implicit state $X(t)$ by simulating a linear continuous-step state space representation in discrete step  \cite{gu2021combining}. The state space representation can be viewed as the inclusion of unobservable variables, providing more information about the system's internal properties \cite{ling2010nonlinear}. Considering the continuous model first:

\begin{align}
    \label{eqn:first_order}
    \dot{X}(t) &= \mathbf{A}X(t) + \mathbf{B}U(t), \\
    \label{eqn:output}
    Y(t) &= \mathbf{C}X(t) + \mathbf{D}U(t),
\end{align}
where $\mathbf{A}$, $\mathbf{B}$, $\mathbf{C}$, and $\mathbf{D}$ are state, control, output, and command matrix. We define $\mathcal{F}(U(t))$ as the states $X(t)$, meaning that the output of \textit{selecting operator} contains implicit information and $\mathbf{T}$ as the output matrix $\mathbf{C}$. Additionally, $\mathcal{H}(U(t))$ can be expressed as the matrix multiplication. Obviously, in \fig \ref{fig:comparison}, each part of CFAR has a correspondence relationship in the linear state space model. After adapting CFAR to the linear state space model, we need to discretize the model so that it can perform iterative training on the computer. We rewrite \eqn \eqref{eqn:first_order} into an ordinary differential equation (ODE) $\dot{X}(t)=f(t,X(t))$ has an equivalent integral equation $X(t) = X(t_0) + \int_{t_0}^{t} f(s,X(s))ds$. 

\textbf{Function approximation}. According to \textit{Picard–Lindelöf theorem} \cite{ince1956ordinary}, given an initial value problem,  we can solve the problem by storing approximation for $X(t)$, and keeping the integral format fixed when iterating given initial function $X_0(t)$ and converge to $X(t)$, defined by:
\begin{align}
    \centering
    X^{(0)}(t) &= X_0(t), \\
    X^{(\ell)}(t) &= X_0(t) + \int_{t_0}^{t} f(s,X^{(\ell-1)}(s))ds,
    \label{eqn:picard_iter}
\end{align}
which means the approximations of the ODE are a sequence of Picard iterates $\{X^{(0)}(t), X^{(1)}(t) \dots\}$. When \eqn \eqref{eqn:picard_iter} in the $\ell$-th iteration, the integral will hold the previous estimate of $X^{(\ell)}(t)$ fixed.

\begin{figure}[t]
    \centering
    \includegraphics[width=0.8\linewidth]{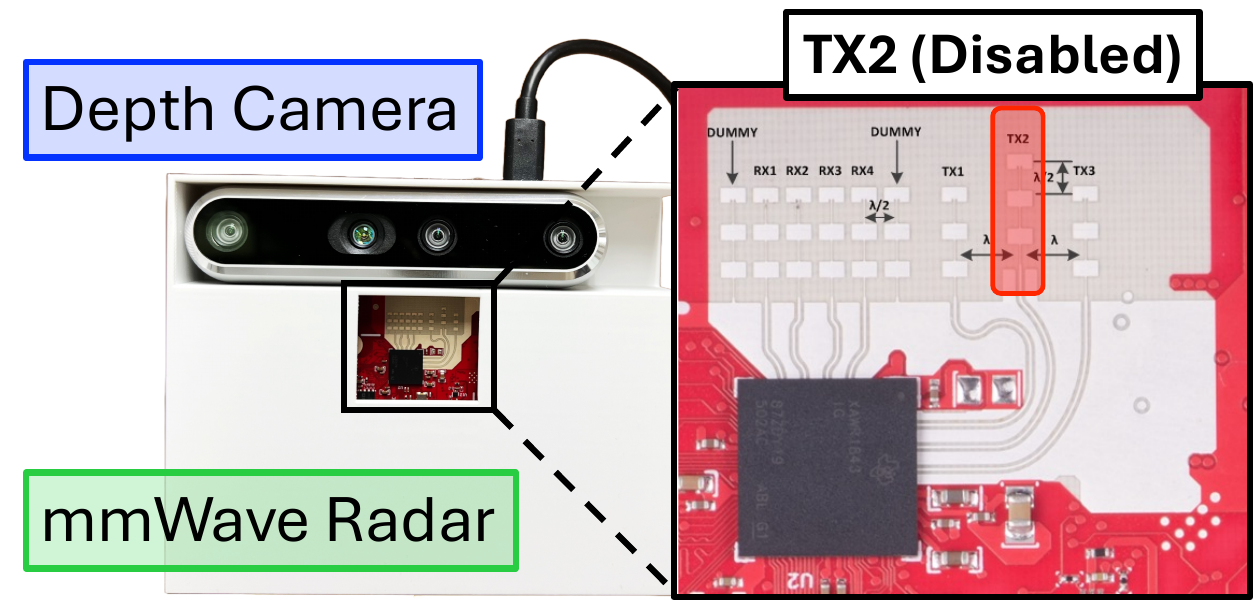}
    \caption{Layout of the sensing node contains a TI FMCW mmWave radar and an Intel RealSense depth camera.}
    \label{fig:node}
    \vspace{-0.1in}
\end{figure}

\textbf{Step approximation and discretization}. Meanwhile, we need to calculate numerical integration on the right-hand side of \eqn \eqref{eqn:picard_iter} for each function approximation iteration, which can be converted to calculate the series of discrete-step values $\{X^{(\ell)}(t_0), X^{(\ell)}(t_1),\dots\}$:
\begin{equation}
    \centering
    \label{eqn:step_iteration}
    X^{(\ell)}(t_{j}) = X^{(\ell)}(t_{j-1}) + \int_{t_{j-1}}^{t_{j}} f(s,X^{(\ell)}(s))ds,
\end{equation}
then we apply the \textit{Trapezoidal rule}\cite{zhang2007performance} to approximate the integral. Given step size $\Delta t_j = t_j - t_{j-1}$:
\begin{equation}
    \begin{aligned}
        \label{eqn:trap}
        X^{(\ell)}(t_{j}) - X^{(\ell)}(t_{j-1}) =
        \int_{t_{j-1}}^{t_{j}} f(s,X^{(\ell)}(s))ds\\
        \approx \frac{\Delta t_j}{2} [f(t_j, X^{(\ell)}(t_j)) + f(t_{j-1}, X^{(\ell)}(t_{j-1}))].
    \end{aligned}
\end{equation}

Considering $f(t,X(t)) = \mathbf{A}X(t) + \mathbf{B}U(t)$, the integral treats $U(t)$ as a fixed value, therefore, we define $U(t_j)=\frac{1}{\Delta t_j} \int_{t_{j-1}}^{t_j} U(s)ds$. Substituting \eqn \eqref{eqn:first_order} into \eqn \eqref{eqn:trap} yields:
\begin{equation}
    \begin{aligned}
        X^{(\ell)}(t_{j}) - X^{(\ell)}(t_{j-1}) = \int_{t_{j-1}}^{t_{j}} (\mathbf{A}X(s) + \mathbf{B}U(s))ds \\
        \approx \frac{\Delta t_j}{2}\mathbf{A}(X^{(\ell)}(t_{j}) + X^{(\ell)}(t_{j-1})) + \Delta t_{j} \mathbf{B}U(t_j),
    \end{aligned}
\end{equation}
which means the discrete $X^{(\ell)}(t_j)$ updates as:
\begin{equation}
    \begin{aligned}
        X^{(\ell)}(t_j) = &(\mathbf{I}-\frac{\Delta t_j}{2} \mathbf{A})^{-1}(\mathbf{I}+\frac{\Delta t_j}{2}  \mathbf{A}) X^{(\ell)}(t_{j-1}) \\
        + &(\mathbf{I}-\frac{\Delta t_j}{2} \mathbf{A})^{-1} \Delta t_j \mathbf{B} U(t_j).
    \end{aligned}
\end{equation}

We redefine $X^{(\ell)}(t_j) = \overline{\mathbf{A}} X^{(\ell)}(t_{j-1}) + \overline{\mathbf{B}} U(t_j)$ and $X^{(\ell)}(t_j)$ and $U(t_j)$ as the discrete sequence notation $X^{(\ell)}[n]$ and $U[n]$, so the Picard iterative discrete linear state model turns:
\begin{align}
    \label{eqn:discrete_first_order}
    X^{(\ell)}[n] &= \overline{\mathbf{A}}X^{(\ell)}[n-1] + \overline{\mathbf{B}}U[n], \\
    \label{eqn:discrete_output}
    Y[n] &= \mathbf{C}X^{(\ell)}[n] + \mathbf{D}U[n],
\end{align}
where $X^{(\ell)}[n]$ converges to $X[n]$ after Picard iteration. The proposed approach can be viewed as a convolution and step size $\Delta t_j$ can control the width of the convolutional kernel, which would be automatically learned when training \cite{gu2021combining}, meaning that the size of convolution window input can be trainable.

\subsection{Training Details}

\begin{figure}[t]
    \centering
    \includegraphics[width=1\linewidth]{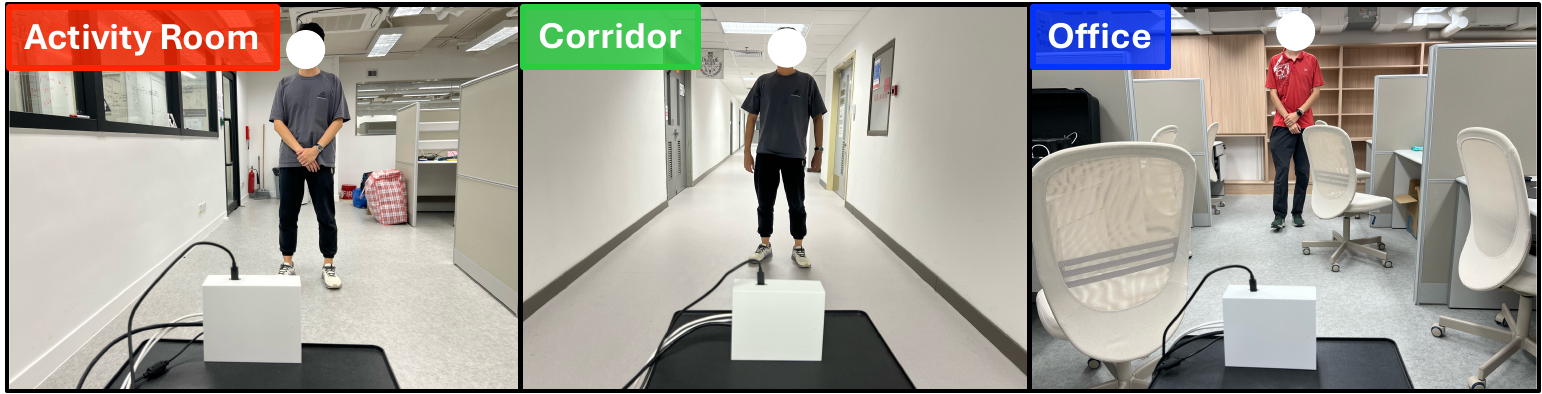}
    \caption{Data collection scenarios include activity room, corridor, and office.}
    \label{fig:scenarios}
\end{figure}

\label{sec:training_details}
The input $U \in \mathbb{R}^{L}$, where $L$ is the length of the flattened input sample and the output of each module is $Y \in \mathbb{R}^{L}$. The state matrix is $X \in \mathbb{R}^{N \times L}$, where $N$ is the feature dimension of the state matrix. Therefore, the other matrices in the model are $\mathbf{A} \in \mathbb{R}^{N \times N}$, $\mathbf{B} \in \mathbb{R}^{N}$, $\mathbf{C} \in \mathbb{R}^{N}$, and $\mathbf{D} \in \mathbb{R}$. Simply, we define the initial state $X[-1] = 0$ and recursively substitute \eqn \eqref{eqn:discrete_first_order} into \eqn \eqref{eqn:discrete_output} yields (omit the Picard iteration number):
\begin{equation}
    Y[n] = \sum_{k=0}^{n} \mathbf{C} (\overline{\mathbf{A}})^{k} \overline{\mathbf{B}}U[n-k] + \mathbf{D}U[n].
\end{equation}

By introducing the \textit{Krylov function} \cite{gu2021combining}:
\begin{equation}
    \mathcal{K}(\mathbf{A},\mathbf{B},\mathbf{C}) = (\mathbf{C}\mathbf{A}^{k}\mathbf{B})_{k\in{[L]}},
\end{equation}
the output $Y$ can be expressed as the non-circular convolution:
\begin{equation}
    Y = \mathcal{K}(\overline{\mathbf{A}}, \overline{\mathbf{B}}, \mathbf{C}) \ast U + \mathbf{D}U,
\end{equation}
where the convolution operation can be efficiently implemented using the Fast Fourier Transform to accelerate computation.

We construct our model by stacking two modules, incorporating linear transformation, layer normalization, and GELU \cite{hendrycks2016gaussian} as the activation function to introduce non-linearity into the network, resulting in approximately 260K parameters. Before training, the input is cloned into $H$ independent copies, enabling the multiple non-interacting training flows simultaneously. By applying $N=256$ and $H=256$, the outputs $Y$ are averaged along the corresponding axis. To enhance training stability during iteration, we apply the translated Legendre (LegT) method to model the state matrices from the HiPPO framework \cite{gu2020hippo}. Additionally, we introduce a residual connection from the input layer to the final layer, followed after a Sigmoid function, to improve the representative capacity of the stacked state space neural network. The model is trained using BCE loss. The model can be seen in \fig \ref{fig:model}.

\begin{figure}[t]
    \centering
    \includegraphics[width=0.85\linewidth]{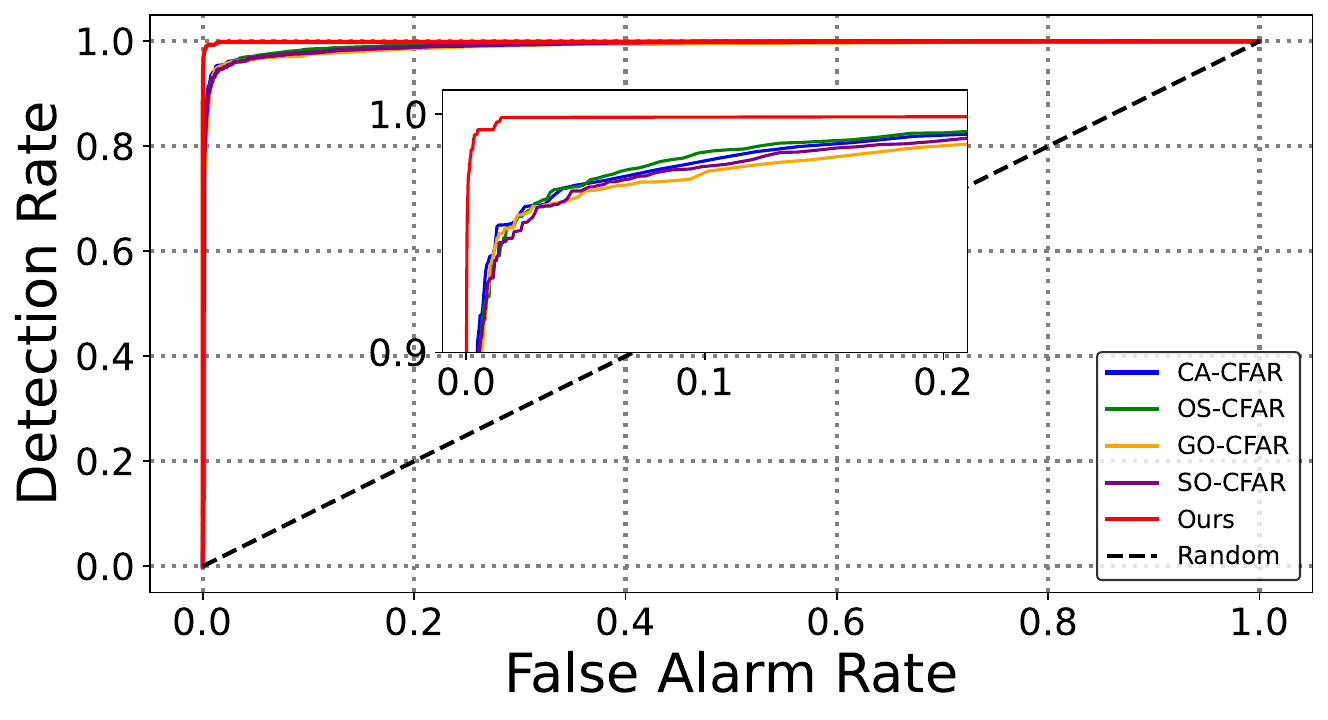}
    \caption{ROC curves of CFAR methods compared to our proposed method.}
    \label{fig:roc_curve}
\end{figure}

\section{Experiments}
\label{sec:experiments}
\subsection{Experiments Setup}
\label{sec:exp_setup}
We focus on 2D range-azimuth data in our experiment. We take the range-azimuth spectrum of TI IWR1843BOOST FMCW mmWave radar as the input and use the point cloud of Intel RealSense D455 depth camera as the ground truth. We encapsulate them in a 3D-printed case to a sensing node for data collection, as illustrated in \fig \ref{fig:node}. The radar has a default azimuth angular resolution of 15$^\circ$, and the frame length of the radar is set to 50 ms. 
The radar's maximum range is around 8.6 m and azimuth FoV is 120$^\circ$. For the depth camera, the detectable depth is from 0.6 m to 6 m and the equivalent azimuth FoV is 87$^\circ$.

As shown in \fig \ref{fig:scenarios}, we consider both single-user and multi-user cases in different scenarios, including an activity room, corridor, and office. In total, we collect 2,222 samples, each including both the radar data and RGB \& depth images of the depth camera. We perform range-FFT with 256 points, beamforming, and calibration on the radar data to calculate the range-azimuth spectrum. To label the ground truth of the spectrum, we utilize YOLO v8 \cite{Jocher_Ultralytics_YOLO_2023} to detect the bounding box of the target, and then employ Segment Anything \cite{kirillov2023segment} to extract the target pixels within the bounding box on the RGB image. We then convert the corresponding pixels of the depth image into the range-azimuth point cloud image, which will be downsampled into a binary mask image (\ie, 1 for occupied cells and 0 otherwise) with the same size as the input spectrum. Due to the range and azimuth differences between the two sensors, we crop them to the overlapped field, resulting in a range of 0.6m to 6m and an azimuth from -43.5° to 43.5°.

\subsection{Results}

We use two metrics to evaluate our system: detection rate ($\mathbf{p}_d$) and false alarm rate ($\mathbf{p}_f$). The detection rate is defined as the ratio of detected cells in the radar output that fall within the target's bounding box determined by YOLOv8 on the RGB image. 
The false alarm rate denotes how many cells are detected outside the bounding box, meaning there is no presence. 

We implement our proposed method and compare it with various CFAR methods and CNN-based \cite{yavuz2021radar} and RNN-based \cite{baird2020cnn} networks for comparison. For the trainable methods (\ie, ours, CNNs, and RNNs), we train all models on the same dataset and evaluate both the detection and false alarm rate. For CFAR methods, we evaluate the detection rate and false alarm rate separately and tune the adjustable parameters of CFAR, ensuring that the other metric is maintained at the same level as in the proposed method. 

\begin{figure}[t]
    \centering
    \begin{minipage}{0.33\linewidth}
        \centering
        \subfloat[Spectrum of single.]{
        \label{subfig:audio_encoder_a}
        \includegraphics[width=\linewidth]{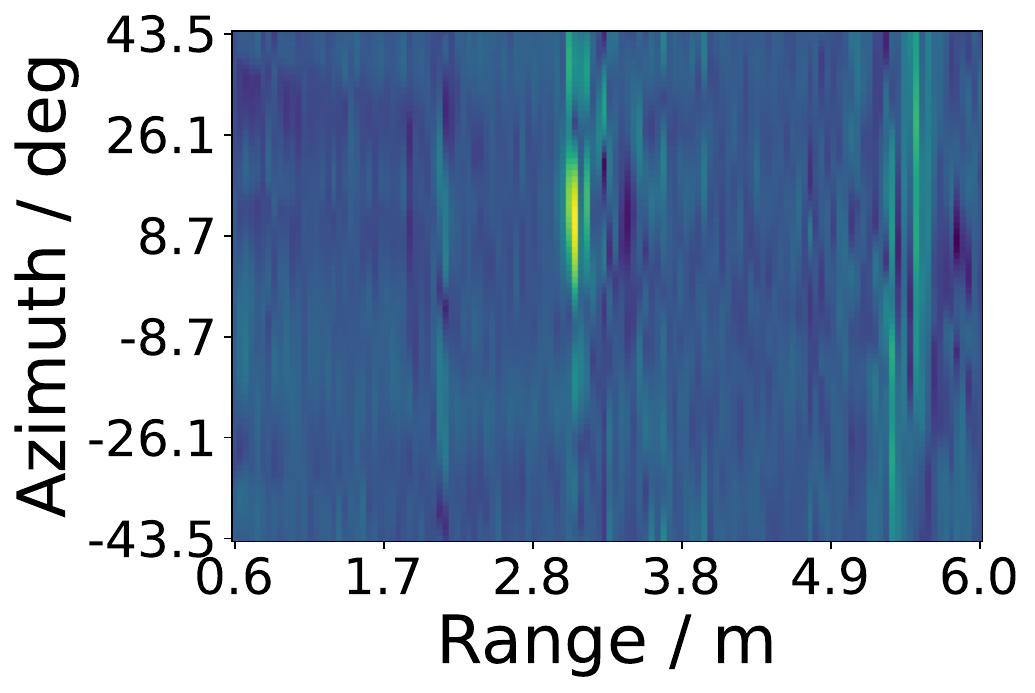}
        }
    \end{minipage}%
    \begin{minipage}{0.33\linewidth}
        \centering
        \subfloat[Ours output.]{
        \label{subfig:video_encoder_b}
        \includegraphics[width=\linewidth]{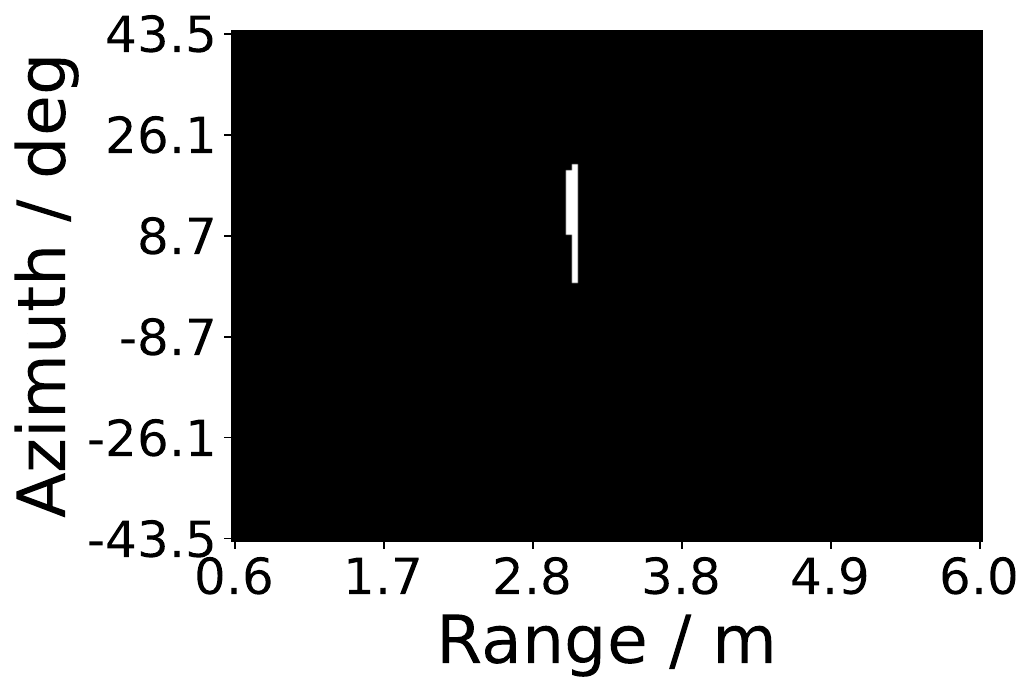}
        }
    \end{minipage}%
    \begin{minipage}{0.33\linewidth}
        \centering
        \subfloat[Point cloud of single.]{
        \label{subfig:video_encoder_c}
        \includegraphics[width=\linewidth]{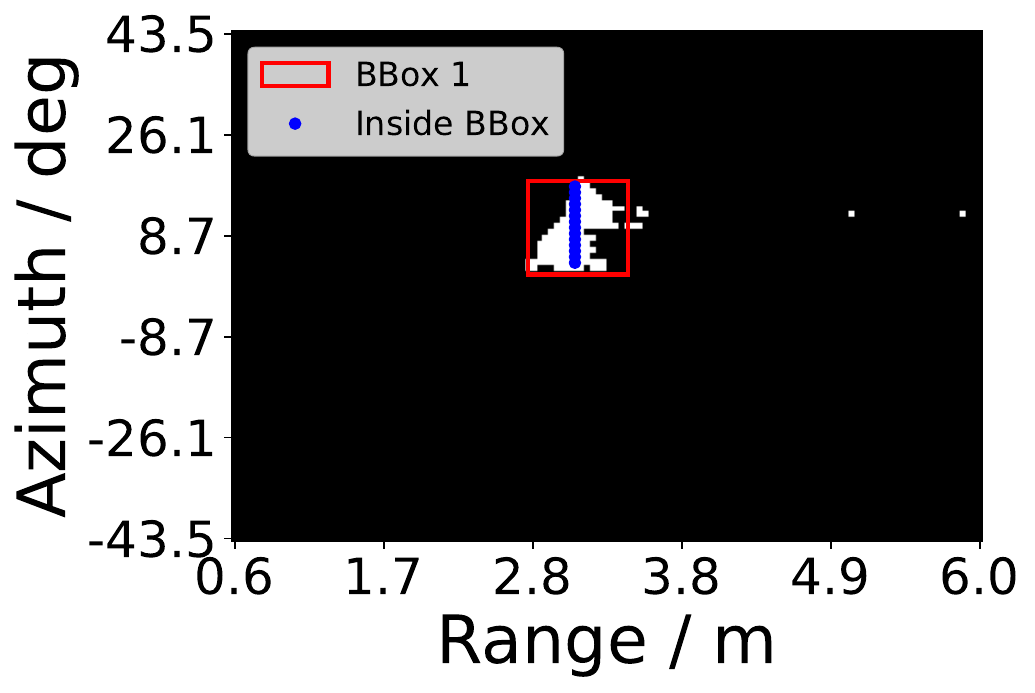}
        }
    \end{minipage}
    \begin{minipage}{0.33\linewidth}
        \centering
        \subfloat[Spectrum of multiples.]{
        \label{subfig:audio_encoder_d}
        \includegraphics[width=\linewidth]{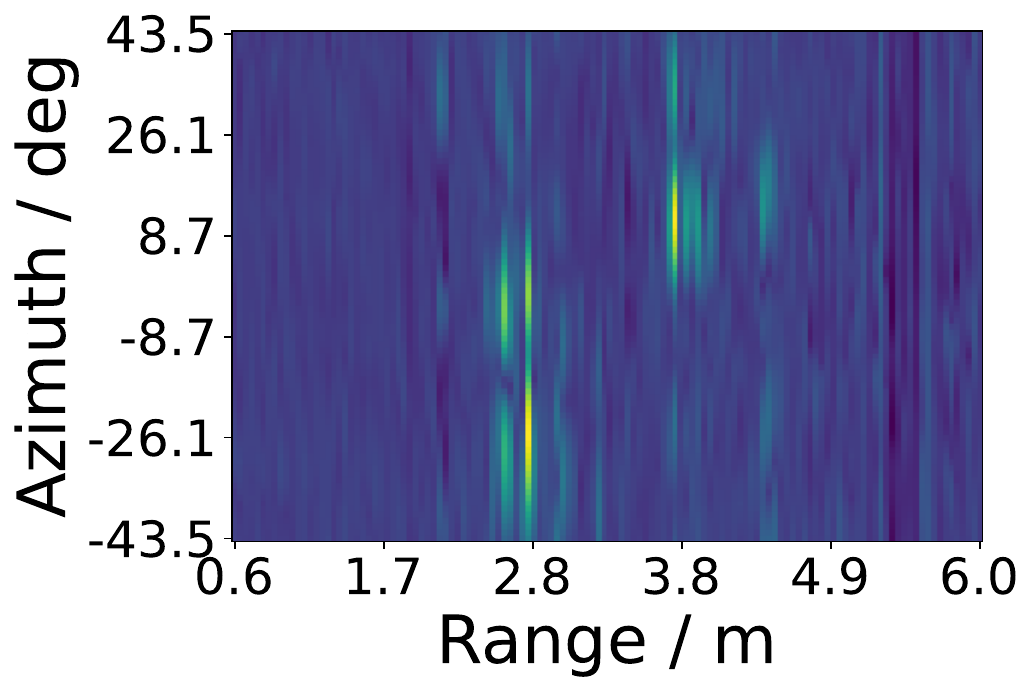}
        }
    \end{minipage}%
    \begin{minipage}{0.33\linewidth}
        \centering
        \subfloat[Ours output.]{
        \label{subfig:video_encoder_e}
        \includegraphics[width=\linewidth]{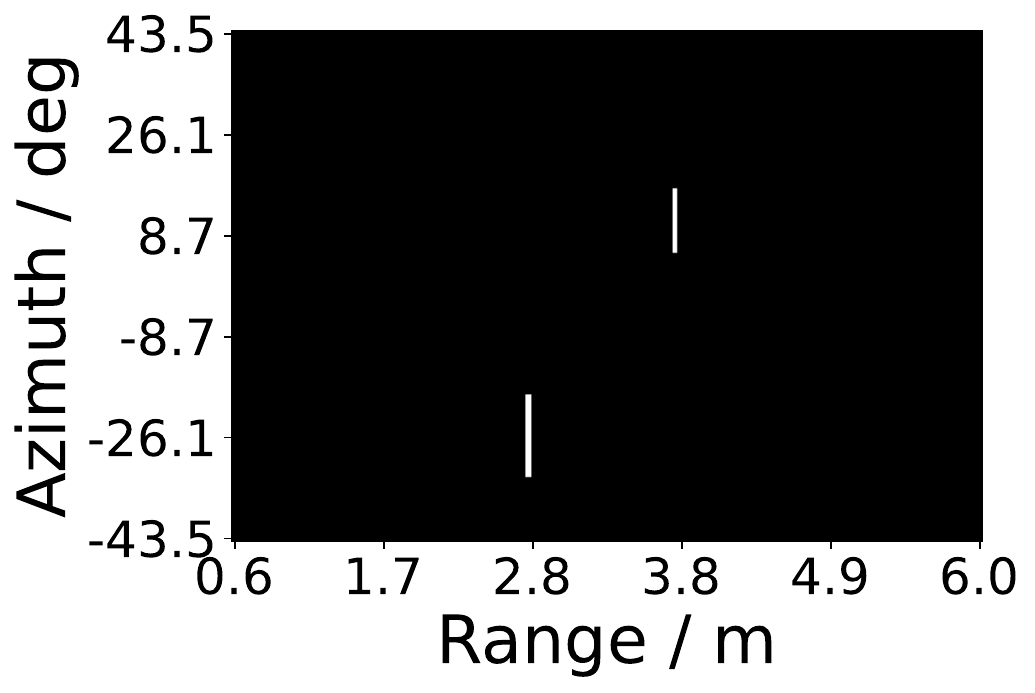}
        }
    \end{minipage}%
    \begin{minipage}{0.33\linewidth}
        \centering
        \subfloat[Point cloud of multiples.]{
        \label{subfig:video_encoder_f}
        \includegraphics[width=\linewidth]{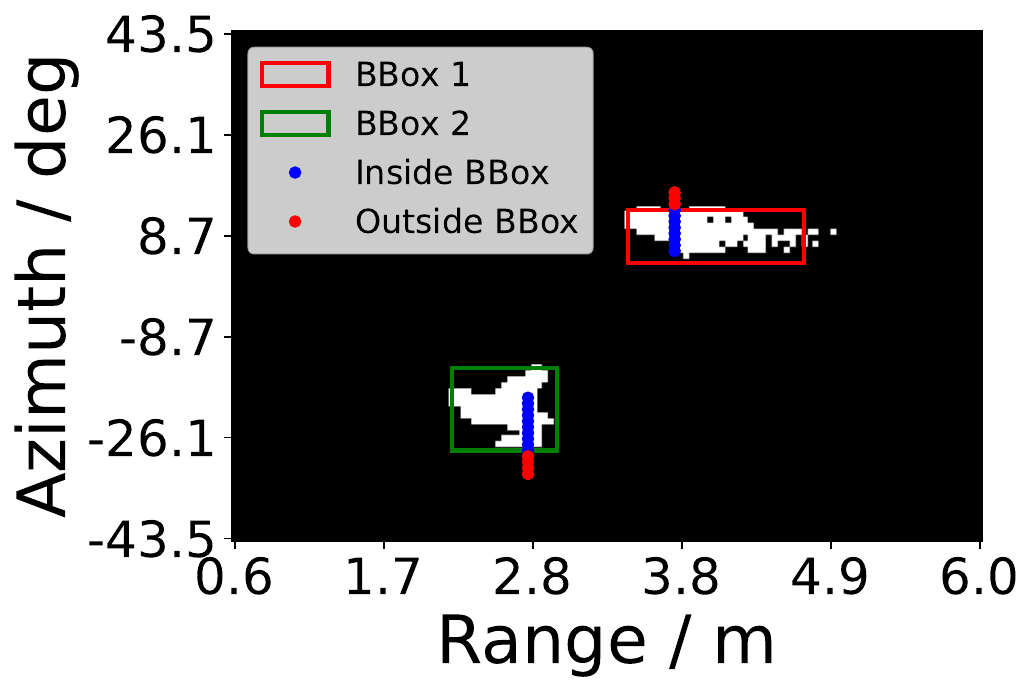}
        }
    \end{minipage}

    \caption{Samples of Results. \rm{Along each row, the figures demonstrate the spectrum of the mmWave radar, the output of our method, and the point cloud of the depth camera.}}
    \label{fig:architecture_encoders}
\end{figure}

\begin{table}[h]
    \centering
    \caption{Experimental results.}
    \label{tab:cell_results}
    
    \begin{tabular}{c|cc}
    \toprule
      \textbf{Methods} & \textbf{$\mathbf{p}_d$} & \textbf{$\mathbf{p}_f$} \\
    \cmidrule{1-3}
      \textbf{Ours} & \textbf{92.60\%} & \textbf{0.06\%} \\
      CNNs \cite{yavuz2021radar} & 88.76\% & 0.17\% \\
      RNNs \cite{baird2020cnn} & 83.94\% & 0.46\% \\
      \cmidrule{1-3}
      \multicolumn{3}{l}{False Alarm Rate Fixed at the Same Level} \\
      \cmidrule{1-3}
      CA-CFAR & 7.80\% & 0.06\%\\
      OS-CFAR & 11.10\% & 0.06\% \\
      GO-CFAR & 6.02\% & 0.06\% \\
      SO-CFAR & 12.62\% & 0.06\% \\
      \cmidrule{1-3}
      \multicolumn{3}{l}{{Detection Rate Fixed at the Same Level}} \\
      \cmidrule{1-3}
      CA-CFAR & 92.09\% & 1.08\% \\
      OS-CFAR & 93.14\% & 1.40\% \\
      GO-CFAR & 92.84\% & 1.51\% \\
      SO-CFAR & 92.55\% & 1.53\% \\      
      \bottomrule
    \end{tabular}
\end{table}

The results are presented in \tab \ref{tab:cell_results}.
As seen, the CNN-based and RNN-based methods, benefiting from prior knowledge, outperform CFAR methods in both detection rate and false alarm rate. However, our approach achieves a higher detection rate and lower false alarm rate than both categories of existing approaches. Compared to CFAR methods, when the false alarm rate is fixed at $0.06\%$, CFAR cannot achieve a higher detection rate. Similarly, when the detection rate is fixed, CFAR's false alarm is, on average, 10 times higher than ours. \fig \ref{fig:roc_curve} demonstrates the ROC curves comparing ours with CFAR methods, clearly showing the superior performance of our method. Additionally, as illustrated in the \tab \ref{tab:scenarios}, we conduct various evaluation experiments to confirm further the robustness and generalization of the proposed method, including the single and multiple targets scenarios and the unseen dataset.
Moreover, our method is significantly more efficient, with only 260K parameters compared to 740K for the CNN-based model and 6M for the RNN-based model. The proposed method also offers better interpretability than these black-box neural networks.

\begin{table*}[htbp!]
    \centering
    \caption{Experimental results on different scenarios, including single and multiple targets, and unseen dataset.}
    \label{tab:scenarios}
    \setlength{\tabcolsep}{4pt} %
    \begin{tabular}{c|cc|cc|cc}
    \toprule
    \multirow{2}{*}{\textbf{Methods}} & \multicolumn{2}{c|}{\textbf{Single}} & \multicolumn{2}{c|}{\textbf{Multiple}} & \multicolumn{2}{c}{\textbf{Unseen}} \\
      & $\mathbf{p}_d$ & $\mathbf{p}_f$ & $\mathbf{p}_d$ & $\mathbf{p}_f$ & $\mathbf{p}_d$ & $\mathbf{p}_f$ \\ 
    \cmidrule(lr){1-1} \cmidrule(lr){2-3} \cmidrule(lr){4-5} \cmidrule(lr){6-7}
    \textbf{Ours} & \textbf{94.72\%} & \textbf{0.04\%} & \textbf{86.89\%} & \textbf{0.09\%} & \textbf{91.92\%} & \textbf{0.06\%} \\ 
    CNNs \cite{yavuz2021radar} & 90.68\%  & 0.17\%  & 73.12\%  & 0.18\%  & 87.09\%  & 0.17\%  \\ 
    RNNs \cite{baird2020cnn} & 82.99\%  & 0.47\%  & 70.73\%  & 0.41\%  & 76.21\%  & 0.46\%  \\ 
    \bottomrule
    \end{tabular}
\end{table*}

\section{Conclusion}
\label{sec:conclusion}
This paper introduces an innovative target detection method that integrates CFAR's classical pipeline with the unfolding state space model, preserving the interpretability of traditional signal processing techniques. Our proposed approach utilizes two stacked modules with only 260K parameters. Our approach outperforms both CNN-based and RNN-based detectors. Compared to CFAR variants, the proposed method shows approximately a 10$\times$ improvement in detection rate at the same false alarm rate and a 10$\times$ reduction in false alarms at the same detection rate. These results highlight our remarkable performance in trainable and interpretable target detection.

\bibliographystyle{IEEEbib}
\bibliography{refs}

\begin{thebibliography}{10}

\bibitem{villeneuve1999improved}
Pierre~V Villeneuve, Herbert~A Fry, James~P Theiler, William~B Clodius, Barham~W Smith, and Alan~D Stocker,
\newblock ``Improved matched-filter detection techniques,''
\newblock in {\em Imaging Spectrometry V}. SPIE, 1999, vol. 3753, pp. 278--285.

\bibitem{ureten1999bayesian}
Oktay Ureten, Nur Serinken, et~al.,
\newblock ``Bayesian detection of radio transmitter turn-on transients.,''
\newblock in {\em NSIp}, 1999, pp. 830--834.

\bibitem{gandhi1989adaptive}
Gandhi and Kassam,
\newblock ``An adaptive order statistic constant false alarm rate detector,''
\newblock in {\em IEEE 1989 International Conference on Systems Engineering}. IEEE, 1989, pp. 85--88.

\bibitem{hansen1973constant}
V~Gregers Hansen,
\newblock ``Constant false alarm rate processing in search radars. in radar—present and future,''
\newblock in {\em IEE Conf Publ}, 1973, vol. 105, p. 325.

\bibitem{trunk1974modified}
GV~Trunk, BH~Cantrell, and FD~Queen,
\newblock ``Modified generalized sign test processor for 2-d radar,''
\newblock {\em IEEE Transactions on Aerospace and Electronic Systems}, , no. 5, pp. 574--582, 1974.

\bibitem{bandiera2007adaptive}
Francesco Bandiera, Antonio De~Maio, and Giuseppe Ricci,
\newblock ``Adaptive cfar radar detection with conic rejection,''
\newblock {\em IEEE Transactions on Signal Processing}, vol. 55, no. 6, pp. 2533--2541, 2007.

\bibitem{khalighi2000adaptive}
Mohammad~Ali Khalighi and Mohammad~Hasan Bastani,
\newblock ``Adaptive cfar processor for nonhomogeneous environments,''
\newblock {\em IEEE Transactions on Aerospace and Electronic Systems}, vol. 36, no. 3, pp. 889--897, 2000.

\bibitem{zhu2021robust}
Xinchao Zhu, Lingying Tu, Shun Zhou, and Zhengwen Zhang,
\newblock ``Robust variability index cfar detector based on bayesian interference control,''
\newblock {\em IEEE Transactions on Geoscience and Remote Sensing}, vol. 60, pp. 1--9, 2021.

\bibitem{wang2017intelligent}
Leiou Wang, Donghui Wang, and Chengpeng Hao,
\newblock ``Intelligent cfar detector based on support vector machine,''
\newblock {\em IEEE Access}, vol. 5, pp. 26965--26972, 2017.

\bibitem{akhtar2018neural}
Jabran Akhtar and Karl~Erik Olsen,
\newblock ``A neural network target detector with partial ca-cfar supervised training,''
\newblock in {\em 2018 International Conference on Radar (RADAR)}. IEEE, 2018, pp. 1--6.

\bibitem{lin2019dl}
Chia-Hung Lin, Yu-Chien Lin, Yue Bai, Wei-Ho Chung, Ta-Sung Lee, and Heikki Huttunen,
\newblock ``Dl-cfar: A novel cfar target detection method based on deep learning,''
\newblock in {\em 2019 IEEE 90th Vehicular Technology Conference (VTC2019-Fall)}. IEEE, 2019, pp. 1--6.

\bibitem{diskin2024cfarnet}
Tzvi Diskin, Yiftach Beer, Uri Okun, and Ami Wiesel,
\newblock ``Cfarnet: deep learning for target detection with constant false alarm rate,''
\newblock {\em Signal Processing}, vol. 223, pp. 109543, 2024.

\bibitem{gu2021combining}
Albert Gu, Isys Johnson, Karan Goel, Khaled Saab, Tri Dao, Atri Rudra, and Christopher R{\'e},
\newblock ``Combining recurrent, convolutional, and continuous-time models with linear state space layers,''
\newblock {\em Advances in neural information processing systems}, vol. 34, pp. 572--585, 2021.

\bibitem{ling2010nonlinear}
Wing-Kuen Ling,
\newblock {\em Nonlinear digital filters: analysis and applications},
\newblock Academic Press, 2010.

\bibitem{ince1956ordinary}
Edward~L Ince,
\newblock {\em Ordinary differential equations},
\newblock Courier Corporation, 1956.

\bibitem{zhang2007performance}
Guofeng Zhang, Tongwen Chen, and Xiang Chen,
\newblock ``Performance recovery in digital implementation of analogue systems,''
\newblock {\em SIAM journal on control and optimization}, vol. 45, no. 6, pp. 2207--2223, 2007.

\bibitem{hendrycks2016gaussian}
Dan Hendrycks and Kevin Gimpel,
\newblock ``Gaussian error linear units (gelus),''
\newblock {\em arXiv preprint arXiv:1606.08415}, 2016.

\bibitem{gu2020hippo}
Albert Gu, Tri Dao, Stefano Ermon, Atri Rudra, and Christopher R{\'e},
\newblock ``Hippo: Recurrent memory with optimal polynomial projections,''
\newblock {\em Advances in neural information processing systems}, vol. 33, pp. 1474--1487, 2020.

\bibitem{Jocher_Ultralytics_YOLO_2023}
Glenn Jocher, Ayush Chaurasia, and Jing Qiu,
\newblock ``{Ultralytics YOLO},'' Jan. 2023.

\bibitem{kirillov2023segment}
Alexander Kirillov, Eric Mintun, Nikhila Ravi, Hanzi Mao, Chloe Rolland, Laura Gustafson, Tete Xiao, Spencer Whitehead, Alexander~C Berg, Wan-Yen Lo, et~al.,
\newblock ``Segment anything,''
\newblock in {\em Proceedings of the IEEE/CVF International Conference on Computer Vision}, 2023, pp. 4015--4026.

\bibitem{yavuz2021radar}
Faruk Yavuz,
\newblock ``Radar target detection with cnn,''
\newblock in {\em 2021 29th European Signal Processing Conference (EUSIPCO)}. IEEE, 2021, pp. 1581--1585.

\bibitem{baird2020cnn}
Zachary Baird, Michael~K Mcdonald, Sreeraman Rajan, and Simon~J Lee,
\newblock ``A cnn-lstm network for augmenting target detection in real maritime wide area surveillance radar data,''
\newblock {\em IEEE Access}, vol. 8, pp. 179281--179294, 2020.

\end{thebibliography}

\end{document}